\documentclass {elsarticle}
\usepackage [utf8] {inputenc}
\usepackage [T1] {fontenc}
\usepackage {url}
\usepackage{changes}

\usepackage{listings}
\usepackage{xcolor}

\definecolor{dkred}{rgb}{0.5,0.1,0}
\definecolor{dkblue}{rgb}{0,0.1,0.5}
\definecolor{ltblue}{rgb}{0,0.5,0.5}
\definecolor{dkgreen}{rgb}{0,0.6,0}
\definecolor{dk2green}{rgb}{0,0.2,0}
\definecolor{dkviolet}{rgb}{0.6,0,0.8}

\newcommand{\code}[1]{\texttt{#1}}

\usepackage {lstcoq}

\begin {document}

\begin {frontmatter}
\title {Formalization of the pumping lemma \\for context-free languages}
\author{Marcus V. M. Ramos}	
\ead{mvmr@cin.ufpe.br}
\address{Centro de Informática, UFPE, Recife, Brazil} 

\author{Ruy J. G. B. de Queiroz}
\ead{ruy@cin.ufpe.br}
\address{Centro de Informática, UFPE, Recife, Brazil} 

\author{Nelma Moreira}
\ead{nam@dcc.fc.up.pt}
\address{Departamento de Ciência de Computadores, Faculdade de Ciências, Universidade do Porto, Porto, Portugal} 

\author{José Carlos Bacelar Almeida}
\ead {jba@di.uminho.pt}
\address{Departamento de Informática, Universidade do Minho, Braga, Portugal} 

\begin {abstract}
Context-free languages (CFLs) are highly important in computer language processing technology as well as in formal language theory. The Pumping Lemma is a property that is valid for all context-free languages, and is used to show the existence of non context-free languages. This paper presents a formalization, using the Coq proof assistant, of the Pumping Lemma for context-free languages.
\end {abstract}

\end {frontmatter}

\section {Introduction}
\label {sec-intro}
The formalization of context-free language theory is key to the certification of compilers and programs, as well as to the development of new languages and tools for certified programming. 

We aim in formalize a substantial part of context-free language theory in the Coq proof assistant, making it possible to reason about it in a fully checked environment, with all the related advantages. Initially, however, the focus has been restricted to context-free grammars and associated results. Pushdown automata and their relation to context-free grammars are not considered at this point.

The work, that started with the formalization of closure properties for context-free grammars \cite {ramos-2014}, evolved later into the formalization of context-free grammar simplification \cite {ramos-2015} and then into the Chomsky normalization of context-free grammars \cite {ramos-2016}. Formalization of simplification enabled the formalization of the Chomsky normalization, which in turn enabled the present formalization of the Pumping Lemma.

In order to follow this paper, the reader is required to have basic knowledge of Coq and of context-free language theory. For the beginner, the recommended starting point for Coq is the book by Bertot and Castéran \cite {bertot-2004}. Background on context-free language theory can be found in \cite {sudkamp-2006} or \cite {ramos-2009}, among others. Previous results, which were used in the formalization of the Pumping Lemma, will not be discussed here and can be retrieved form the above references.

The statement and applications of the Pumping Lemma for CFLs (or Pumping Lemma for short) are presented in Section \ref {sec-pumping}. A typical informal proof, which served as the basis for the present formalization, is described in Section \ref {sec-informal}. Section \ref {sec-background} introduces results obtained previously and which are required for this work. The formalization is then described in Section \ref {sec-formal}, where the definitions and auxiliary lemmas used are discussed in some detail, as well as the Pumping Lemma itself. Section \ref {sec-related} discusses related work by various other researchers and final conclusions are presented in Section \ref {sec-conclusions}.

As far as the authors are aware of, this is the first formalization of the Pumping Lemma for context-free languages in any proof assistant. All the definitions and proof scripts discussed in this paper were written in plain Coq and are available for download at: \\
\url {https://github.com/mvmramos/pumping} \\

\section {Statement and Application}
\label {sec-pumping}

A language is a set of words defined over an alphabet. A context-free grammar is a grammar whose rules have the form $X \rightarrow \beta$, where $X$ is a non-terminal symbol and $\beta$ is a sequence (possibly empty) of terminal and non-terminal symbols. A context-free language is a language that is generated by some context-free grammar. The Pumping Lemma is a property that is verified for all CFLs and was stated and proved for the first time by Bar-Hillel, Perles and Shamir in 1961 \cite {bar-hillel-1961}.

The Pumping Lemma does not characterize the CFLs, however, since it is also verified by some non CFLs. It states that, for every context-free language and for every sentence of such a language that has a certain minimum length, it is possible to obtain an infinite number of new sentences that must also belong to the language. This minimum length depends only on the language defined. In other words (let $\mathcal {L}$ be defined over alphabet $\Sigma$):

$$\forall \mathcal {L}, (\mbox {cfl } \mathcal {L}) \rightarrow \exists n\, |$$
$$\forall \alpha, (\alpha \in \mathcal {L}) \land (|\alpha| \ge n )\rightarrow$$
$$\exists u, v, w, x, y \in \Sigma^*\, |\, (\alpha = uvwxy) \land (|vx| \ge 1) \land (|vwx| \le n)\,\land$$
$$\forall i, uv^iwx^iy \in \mathcal {L}$$

A typical use of the Pumping Lemma is to show that a certain language is not context-free by using the contrapositive of the statement of the lemma. The proof proceeds by contraposition: the language is assumed to be context-free, and this leads to a contradiction from which one concludes that the language in question cannot be context-free. 

As an example, consider the language $\mathcal {L} = \{a^ib^ic^i\,|\,i \ge 1\}$. This language is defined over the alphabet $\{a, b, c\}$ and includes sentences such as $abc, aabbcc, aaabbbccc$.

Should $\mathcal {L}$ be context-free, then the Pumping Lemma should hold for it. Consider $n$ to be the constant of the Pumping Lemma and let's choose the sentence $\alpha = a^nb^nc^n$. Clearly, $\alpha \in \mathcal {L}$ and $|\alpha| = 3*n \ge n$. Thus, $\alpha = uvwxy$ such that $|vx| \ge 1$, $|vwx| \le n$ and $uv^iwx^iy \in \mathcal {L}, i \ge 0$. 

However, it is easy to observe that, due to its length limitation, the sentence $vwx$ should contain only one or two different symbols (namely, $vwx$ should belong to either $a^*, b^*, c^*, a^*b^*$ or $b^*c^*$). This implies that the repetition of $v$ and $x$ in $uv^iwx^iy$ should increase (or decrease) the number of at most two different symbols while keeping the number of the third symbol unchanged. As a result, the new sentence cannot belong to the language and this proves that the initial hypothesis cannot be true. Thus, $\mathcal {L}$ is not a context-free language.

\section {Informal Proof}
\label {sec-informal}

In short, the Pumping Lemma derives from the fact that the number of non-terminal symbols in the grammar $G$ that generates $\mathcal {L}$ is finite. The classical proof considers that $G$ is in the Chomsky Normal Form (a form in which the rules of the grammar have at most two symbols in the right-hand side), which means that derivation trees have the simpler form of binary trees. Then, if the sentence has a certain minimum length, the frontier of the derivation tree should have two or more instances of the same non-terminal symbol in some path that starts in the root of this tree. Finally, the context-free character of $G$ guarantees that the subtrees related to these duplicated non-terminal symbols can be cut and pasted in such a way that an infinite number of new derivation trees are obtained, each of which is related to a new sentence of the language.

The proof comprises the following steps (more details can be found in \cite {sudkamp-2006} or \cite {ramos-2009}):

\begin {enumerate}
\item Since $\mathcal {L}$ is declared to be a context-free language (predicate \texttt {cfl}), then there exists a context-free grammar $G$ such that $L(G) = \mathcal {L}$;
\item Obtain $G'$ such that $G'$ is in Chomsky Normal Form and $L(G')=L(G)$;
\item Take $n$ as $2^k$, where $k$ is the number of non-terminal symbols in $G'$;
\item Choose $\alpha$ such that $\alpha \in \mathcal {L}$ and $|\alpha| \ge n$;
\item Obtain a derivation tree $t$ that represents the derivation of $\alpha$ in $G'$;
\item Take a path that starts in the root of $t$ and whose length is the height of $t$ plus 1 (maximum length);
\item Then, the height of $t$ should be greater or equal than $k + 1$;
\item This means that the selected path has at least $k + 2$ symbols, being at least $k + 1$ non-terminals and one (the last) a terminal symbol;
\item Since $G'$ has only $k$ non-terminal symbols, this means that this path has at least one non-terminal symbol that appears at least two times in it;
\item Name the duplicated symbols $n_1$ and $n_2$ ($n_1 = n_2$) and the corresponding subtrees $t_1$ and $t_2$ (note that $t_2$ is a subtree of $t_1$ and $t_1$ is a subtree of $t$);
\item It is then possible to prove that the height of $t_1$ is greater than or equal to 2, and less than or equal to $2^k$;
\item Also, that the height of $t_2$  is greater than or equal to 1 and less than or equal to $2^{k-1}$;
\item This implies that the frontier of $t$ can be split into five parts: $u, v, w, x, y$, where $w$ is the frontier of $t_2$ and $vwx$ is the frontier of $t_1$;
\item As a consequence of the heights of the corresponding subtrees, it can be shown that $|vx| \ge 1$ and $|vwx| \le n$;
\item If $t_1$ is removed from $t$, and $t_2$ is inserted in its place, then we have a new tree $t^0$ that represents the derivation of string $uv^0wx^0y = uwy$;
\item If, instead, $t_1$ is inserted in the place where $t_2$ lies originally, then we have a tree $t^2$ that represents the derivation of string $uv^2wx^2y$;
\item Repetition of the previous step generates all trees $t^i$ that represent the derivation of the string $uv^iwx^iy$, $\forall i \ge 2$.
\end {enumerate}

\section {Background}
\label {sec-background}

Formalization in the Coq proof assistant requires the formalization of the Chomsky Normal Form (CNF) for context-free grammars. This, in turn, demands the formalization of context-free grammar simplification (useless and inaccessible symbol elimination and unit and empty rules elimination). For details on how these have been accomplished, please refer to \cite {ramos-2015} and \cite {ramos-2016}. 

The Chomsky Normal Form (CNF) theorem asserts 
$$\forall\,G=(V,\Sigma,P,S),\;\exists\,G'=(V',\Sigma,P',S')\;|$$
$$L(G)=L(G') \land$$
$$\forall\,(\alpha \rightarrow_{G'} \beta) \in P', (\beta \in \Sigma) \lor (\beta \in N\cdot N)$$

That is, every context-free grammar can be converted to an equivalent one whose rules have only one terminal symbol or two non-terminal symbols in the right-hand side. Naturally, this is valid only if $G$ does not generate the empty string. If this is the case, then the grammar that has this format, plus a single rule $S' \rightarrow_{G'} \epsilon$, is also considered to be in the Chomsky Normal Form, and generates the original language, including the empty string. It can also be assured that in either case the start symbol of $G'$ does not appear on the right-hand side of any rule of $G'$.

The CNF theorem has been stated in our formalization as:

\begin{coq}
Theorem g_cnf:
forall g: cfg non_terminal terminal,
(produces_empty g \/ ~ produces_empty g) /\ 
(produces_non_empty g \/ ~ produces_non_empty g) ->
exists g': cfg non_terminal' terminal, 
g_equiv g' g /\ 
(is_cnf g' \/ is_cnf_with_empty_rule g').
\end{coq}

Context-free grammars are represented by record \texttt {cfg} in a way that resembles the definition:

\begin{coq}
Record cfg (non_terminal terminal:Type): Type:= {
start_symbol: non_terminal;
rules: non_terminal -> list (non_terminal + terminal) -> Prop;
rules_finite:
         exists n: nat,
         exists ntl: list non_terminal,
         exists tl: list terminal,
         rules_finite_def start_symbol rules n ntl tl}.
\end{coq}

The predicate \texttt {rules\_finite\_def} assures that the set of rules of the grammar is finite by proving that the length of right-hand side of every rule is equal or less than a given value, and also that both left and right-hand side of the rules are built from finite sets of, respectively, non-terminal and terminal symbols (represented here by lists).

The predicate \texttt {produces g s} asserts that context-free grammar \texttt {g} produces the list of terminals \texttt {s} as a sentence of the language. It is based on the more fundamental notion of \emph {derivation}, present in the whole formalization and defined as:

\begin{coq}
Inductive derives (g: cfg): sf -> sf -> Prop :=
| derives_refl: 
         forall s: sf,
         derives g s s
| derives_step: 
         forall s1 s2 s3: sf,
         forall left: non_terminal,
         forall right: sf,
         derives g s1 (s2 ++ inl left :: s3) ->
         rules g left right ->
         derives g s1 (s2 ++ right ++ s3).
\end{coq}

The predicates used in theorem \texttt {g\_cnf\_final} above assert that:

\begin {itemize}
\item a grammar produces the empty string:
      \begin{coq}
	  Definition produces_empty 
	  (g: cfg non_terminal terminal): Prop:=
	  produces g [].
      \end{coq}
\item a grammar produces a non-empty string:
      \begin{coq}
	  Definition produces_non_empty 
	  (g: cfg non_terminal terminal): Prop:=
	  exists s: sentence, produces g s /\ s <> [].     
	  \end{coq}
\item two grammars are equivalent:
      \begin{coq}
	  Definition g_equiv 
	  (non_terminal non_terminal' terminal: Type) 
	  (g1: cfg non_terminal terminal) 
	  (g2: cfg non_terminal' terminal): Prop:=
      forall s: sentence,
      produces g1 s <-> produces g2 s.
	  \end{coq}
\item a rule is in the Chomsky Normal Form:
      \begin{coq}
	  Definition is_cnf_rule 
	  (left: non_terminal) (right: sf): Prop:=
      (exists s1 s2: non_terminal, right = [inl s1; inl s2]) \/
      (exists t: terminal, right = [inr t]).
	  \end{coq}
\item a grammar is in the Chomsky Normal Form:
      \begin{coq}
	  Definition is_cnf 
	  (g: cfg non_terminal terminal): Prop:=
	  forall left: non_terminal,
	  forall right: sf,
	  rules g left right -> is_cnf_rule left right.
	  \end{coq}
\item a grammar is in the Chomsky Normal Form and has a single empty rule with the start symbol in the left-hand side:
      \begin{coq}
	  Definition is_cnf_with_empty_rule 
	  (g: cfg non_terminal terminal): Prop:=
	  forall left: non_terminal,
	  forall right: sf,
	  rules g left right ->
	  (left = (start_symbol g) /\ right = []) \/
	  is_cnf_rule left right.	  
	  \end{coq}
\end {itemize}

\section {Formalization}
\label {sec-formal}

The formalization follows closely the steps described in Section \ref {sec-informal}. The Pumping Lemma has been stated as follows:

\begin{coq}
Lemma pumping_lemma:
forall l: lang terminal,
(contains_empty l \/ ~ contains_empty l) /\ 
(contains_non_empty l \/ ~ contains_non_empty l) ->
cfl l ->
exists n: nat, 
forall s: sentence, 
l s -> 
length s >= n ->
exists u v w x y: sentence, 
s = u ++ v ++ w ++ x ++ y /\
length (v ++ x) >= 1 /\
length (v ++ w ++ x) <= n /\
forall i: nat, l (u ++ (iter v i) ++ w ++ (iter x i) ++ y).
\end{coq}

A language is defined as a function that maps a sentence (a list of terminal symbols) to a proposition (\texttt {Prop}):

\begin{coq}
Definition lang (terminal Type):= list terminal -> Prop.
\end{coq}

Two languages are equal if they have the same sentences:

\begin{coq}
Definition lang_eq (l k: lang) := 
forall w, l w <-> k w.
\end{coq}

Finally, a language is context-free if it is generated by some context-free grammar:

\begin{coq}
Definition cfl (terminal: Type) (l: lang terminal): Prop:=
exists non_terminal: Type, 
exists g: cfg non_terminal terminal, 
lang_eq l (lang_of_g g).
\end{coq}

\noindent
where \texttt {lang\_of\_g} represents the language generated by grammar \texttt {g}:

\begin{coq}
Definition lang_of_g (g: cfg non_terminal terminal): lang :=
fun w: sentence => produces g w.
\end{coq}

Predicates \texttt {contains\_empty} and \texttt {contains\_non\_empty} are language counterparts of the previously presented grammar predicates \texttt {produces\_empty} and \texttt {produces\_non\_empty}, respectively. Application \texttt {iter l i} on a list \texttt {l} and a natural \texttt {i} yields list \texttt {l\textsuperscript{i}}.

Initially, the type \texttt {btree} (for binary trees) has been defined with the objective of representing derivation trees for strings generated by context-free grammars in Chomsky Normal Form:

\begin{coq}
Inductive btree (non_terminal terminal: Type): Type:=
| bnode_1: non_terminal -> terminal -> btree
| bnode_2: non_terminal -> btree -> btree -> btree.
\end{coq}

The constructors of \texttt {btree} relate to the two possible forms that the rules of a CNF grammar can assume (namely with one terminal symbol or two non-terminal symbols in the right-hand side).

The proof of the Pumping Lemma starts by finding a grammar $G$ that generates the input language $L$ (this is a direct consequence of the predicate \texttt {is\_cfl} and corresponds to step 1 of Section \ref {sec-informal}). Next, we obtain a CNF grammar $G'$ that is equivalent to $G$ (step 2), using previous results. Then, $G$ is substituted for $G'$ and the value for $n$ is defined as $2^k$ (step 3) where $k$ is the length of the list of non-terminals of $G'$ (which in turn is obtained from the predicate \texttt {rules\_finite}).

Lemma \texttt {derives\_g\_cnf\_equiv\_btree} is then applied in order to obtain a \texttt {btree} $t$ that represents the derivation of $\alpha$ in $G'$ (step 5). This lemma is general enough in order to accept that the input grammar might either be a CNF grammar, or a CNF grammar with an empty rule. If this is the case, then we have to ensure that $\alpha \neq \epsilon$, which is true since by assumption $|\alpha| \ge 2^k$. The proof of \texttt {derives\_g\_cnf\_equiv\_btree} is reasonably long and uses induction on the number of derivation steps in $G'$ in order to generate $\alpha$:

\begin{coq}
Lemma derives_g_cnf_equiv_btree:
forall g: cfg non_terminal' terminal,
forall n: non_terminal',
forall s: sentence,
s <> [] ->
(is_cnf g \/ is_cnf_with_empty_rule g) ->
start_symbol_not_in_rhs g ->
derives g [inl n] (map term_lift' s) ->
exists t: btree non_terminal' terminal,
btree_cnf g t /\
broot t = n /\
bfrontier t = s.
\end{coq}

The next step is to obtain a path (a sequence of non-terminal symbols ended by a terminal symbol) that has maximum length, that is, whose length is equal to the height of $t$ plus $1$ (steps 6 and 7). This is accomplished by means of the definition \texttt {bpath} and the lemma \texttt {btree\_ex\_bpath}:

\begin{coq}
Inductive bpath (bt: btree): sf -> Prop:=
| bp_1: forall n: non_terminal,
        forall t: terminal,
        bt = (bnode_1 n t) -> bpath bt [inl n; inr t]
| bp_l: forall n: non_terminal,
        forall bt1 bt2: btree,
        forall p1: sf,
        bt = bnode_2 n bt1 bt2 -> bpath bt1 p1 -> bpath bt ((inl n) :: p1)
| bp_r: forall n: non_terminal,
        forall bt1 bt2: btree,
        forall p2: sf,
        bt = bnode_2 n bt1 bt2 -> bpath bt2 p2 -> bpath bt ((inl n) :: p2).
\end{coq}

\begin{coq}
Lemma btree_ex_bpath:
forall bt: btree,
forall ntl: list non_terminal,
bheight bt >= length ntl + 1 ->
bnts bt ntl ->
exists z: sf,
bpath bt z /\
length z = bheight bt + 1 /\
exists u r: sf,
exists t: terminal,
z = u ++ r ++ [inr t] /\
length u >= 0 /\
length r = length ntl + 1 /\
(forall s: symbol, In s (u ++ r) -> In s (map inl ntl)).
\end{coq}

The length of this path (which is $\ge k + 2$) allows one to infer that it must contain at least one non-terminal symbol that appears at least twice in it (steps 8, 9 and 10). This result comes from the application of the lemma \texttt {pigeon} which represents a list version of the well-known pigeonhole principle:

\begin{coq}
Lemma pigeon:
forall A: Type,
forall x y: list A,
(forall e: A, In e x -> In e y) ->
length x = length y + 1->
exists d: A,
exists x1 x2 x3: list A,
x = x1 ++ [d] ++ x2 ++ [d] ++ x3.
\end{coq}

Since a path is not unique in a tree, it is necessary to use some some other representation that can describe this path uniquely, which is done by the predicate \texttt {bcode} and the lemma \texttt {bpath\_ex\_bcode}:

\begin{coq}
Inductive bcode (bt: btree): list bool -> Prop:=
| bcode_0: forall n: non_terminal,
           forall t: terminal,
           bt = (bnode_1 n t) -> bcode bt []
| bcode_1: forall n: non_terminal,
           forall bt1 bt2: btree,
           forall c1: list bool,
           bt = bnode_2 n bt1 bt2 -> bcode bt1 c1 -> bcode bt (false :: c1)
| bcode_2: forall n: non_terminal,
           forall bt1 bt2: btree,
           forall c2: list bool,
           bt = bnode_2 n bt1 bt2 -> bcode bt2 c2 -> bcode bt (true :: c2).
\end{coq}

The predicate \texttt {bcode} uses a sequence of boolean values (\texttt {false, true}) to respectively select the left or right subtrees in a tree, and thus define a path in it.

\begin{coq}
Lemma bpath_ex_bcode:
forall t: btree,
forall p: sf,
bpath t p -> 
exists c: list bool,
bcode t c /\
bpath_bcode t p c.
\end{coq}

The predicate \texttt {bpath\_bcode} merely ensures that \texttt {bcode c} is valid for \texttt {bpath p} in tree \texttt {t}. Once the path has been identified with a repeated non-terminal symbol, and a corresponding \texttt {bcode} has been assigned to it, lemma \texttt {bcode\_split} is applied twice in order to obtain the two subtrees $t_1$ and $t_2$ that are associated respectively to the first and second repeated non-terminals of $t$. This lemma, which is key in the formalization, has a statement with a number of hypothesis and conclusions which give many useful informations on the newly identified subtree. Among them, its height and its frontier (this one embedded in the definition \texttt {btree\_decompose}):

\begin{coq}
Lemma bcode_split:
forall t: btree,
forall p1 p2: sf,
forall c: list bool,
bpath_bcode t (p1 ++ p2) c ->
length p1 > 0 ->
length p2 > 1 ->
bheight t = length p1 + length p2 - 1 ->
exists c1 c2: list bool,
c = c1 ++ c2 /\
length c1 = length p1 /\
exists t2: btree,
exists x y: sentence,
bpath_bcode t2 p2 c2 /\
btree_decompose t c1 = Some (x, t2, y) /\
bheight t2 = length p2 - 1.
\end{coq}

Function \texttt {btree\_decompose} takes as arguments a tree and a sequence of boolean values, and returns a triple consisting of the subtree located in this position and the two sentences to the left and right of it. It is used to enable reasoning on the frontiers of the subtrees obtained before.

From this information it is then possible to extract most of the results needed to prove the goal (steps 11, 12, 13 and 14), except for the pumping condition. This has been obtained by an auxiliary lemma \texttt {pumping\_aux}, which takes as hypothesis the fact that a tree $t_1$ (with frontier $vwx$) has a subtree $t_2$ (with frontier $w$), both with the same roots, and asserts the existence of an infinite number of new trees obtained by repeated substitution of $t_2$ by $t_1$ or simply $t_1$ by $t_2$, with respectively frontiers $v^iwx^i, i \ge 1$ and $w$, or simply $v^iwx^i, i \ge 0$:

\begin{coq}
Lemma pumping_aux:
forall g: cfg _ _,
forall t1 t2: btree (non_terminal' non_terminal terminal) _,
forall n: _,
forall c1 c2: list bool,
forall v x: sentence,
btree_decompose t1 c1 = Some (v, t2, x) ->
btree_cnf g t1 ->
broot t1 = n ->
bcode t1 (c1 ++ c2) ->
c1 <> [] ->
broot t2 = n ->
bcode t2 c2 ->
(forall i: nat,
 exists t': btree _ _,
 btree_cnf g t' /\
 broot t' = n /\
 btree_decompose t' (iter c1 i) = Some (iter v i, t2, iter x i) /\
 bcode t' (iter c1 i ++ c2) /\
 get_nt_btree (iter c1 i) t' = Some n).
\end{coq}

The proof continues by showing that each of these new trees can be combined with tree $t$ obtained before, thus representing strings $uv^iwx^iy, i \ge 0$ as necessary (steps 15 and 16). 

Finally, it must be proved that each of these trees is related to a derivation in $G'$, which is accomplished by lemma \texttt {btree\_equiv\_produces\_g\_cnf}, the dual version of lemma \texttt {derives\_g\_cnf\_equiv\_btree} (step 17).

The Pumping Lemma has some 400 lines of Coq script, which adds to auxiliary lemmas and an extensive library of lemmas on binary trees and on the relation of binary trees to Chomsky Normal Form grammars. The whole approach is constructive, except for the proof of the \texttt {pigeon} lemma, which was formalized with classical logic extensions.

\section {Related Work}
\label {sec-related}
Context-free language theory formalization is a relatively new area of research, with some results already obtained with a diversity of proof assistants, including Coq, HOL4 and Agda. Most of the effort started in 2010 and have been devoted to the certification and validation of parser generators. Examples of this are the works of Koprowski and Binsztok (using Coq, \cite {koprowski-2010}), Ridge (using HOL4, \cite {ridge-2011}), Jourdan, Pottier and Leroy (using Coq, \cite {jourdan-2012}) and, more recently, Firsov and Uustalu (in Agda, \cite {firsov-2014}). 

On the more theoretical side, on which the present work should be considered, Norrish and Barthwal published on general context-free language theory formalization using the using HOL4 proof assistant \cite {barthwal-norrish-2010a,barthwal-norrish-2010b,barthwal-norrish-2013}, including the existence of normal forms for grammars, pushdown automata and closure properties. Recently, Firsov and Uustalu proved the existence of a Chomsky Normal Form grammar for every general context-free grammar, using the Agda proof assistant \cite {firsov-2015}.

A special case of the Pumping Lemma, namely the Pumping Lemma for regular languages, is included in a comprehensive work on the formalization of regular languages \cite {doczkal-2013} using SSRreflect, an extension of Coq.

\section {Conclusions}
\label {sec-conclusions}

The formalization of the Pumping Lemma for context-free languages represents the culmination of an effort that started with closure properties for context-free grammars \cite {ramos-2014} and continued with simplification for context-free grammars \cite {ramos-2015} and the Chomsky Normal Form \cite {ramos-2016}. The whole formalization has 20.000+ lines of Coq script and was developed over a period of two years.

The Pumping Lemma is a significant result in language theory in general and this is, as far as the authors are aware of, the first formalization ever of it, 54 years after is was stated and proved for the first time. It has to be seen against the backdrop of the important and well sought after goal of formalizing fundamental results in language theory, as well as formalizing mathematics in general.

The libraries developed to support this formalization will hopefully play an equally important role, as they include general results on context-free language theory that can be used or adapted to prove other results. The whole work can serve to different purposes, including the continued formalization of language theory and the teaching of formal languages, formalization and Coq itself.

\section*{Bibliography}
\bibliographystyle {elsarticle-harv}
\bibliography {article}
\end {document}